# Laser Guide Star Adaptive Optics without Tip-tilt

Richard Davies[1], Sebastian Rabien[1], Chris Lidman[2], Miska Le Louarn[2], Markus Kasper[2], Natascha M. Förster Schreiber[1], Veronica Roccatagliata[3], Nancy Ageorges[1], Paola Amico[2], Christoph Dumas[2], Filiippo Mannucci[4]

[1] Max-Planck-Institut für extraterrestrische Physik, Garching, Germany
[2] ESO
[3] Max-Planck-Institut für Astronomie, Heidelberg, Germany
[4] INAF-Istituto di Radioastronomia, Firenze, Italy

Adaptive optics (AO) systems allow a telescope to reach its diffraction limit at near infrared wavelengths. But to achieve this, a bright natural guide star (NGS) is needed for the wavefront sensing, severely limiting the fraction of the sky over which AO can be used. To some extent this can be overcome with a laser guide star (LGS). While the laser can be pointed anywhere in the sky, one still needs to have a natural star, albeit fainter, reasonably close to correct the image motion (tip-tilt) to which laser guide stars are insensitive. There are in fact many astronomical targets without suitable tip-tilt stars, but for which the enhanced resolution obtained with the Laser Guide Star Facility (LGSF) would still be very beneficial. This article explores what adaptive optics performance one might expect if one dispenses with the tip-tilt star, and in what situations this mode of observing might be needed.

## To Find a Star, or not

The constraints for adaptive optics with a natural guide star mean that very few astronomical targets are suited to this technique: to get the best performance it has to be brighter than V~13mag and within 15-20". With a laser guide star these restrictions are very much relaxed; but even with a LGS, one still needs to find a natural guide star for the tip-tilt correction. At the VLT this tip-tilt star must be brighter than V~17mag within about 60". While this vastly increases the number of targets to which adaptive optics can be applied, there are still important cases that slip through the net. Perhaps the most obvious of these are the so-called "deep fields" (such as the GOODS-CDFS) which are currently extensively surveyed at all accessible wavelengths to study galaxy formation and evolution out to very high redshift. But galaxies at high redshift are typically 1 arcsec or less across, so such work would benefit enormously from adaptive optics techniques that enable the galaxies to be resolved. Yet these fields are barely accessible to AO because one of their prime selection criteria is to contain as few bright stars as possible to avoid saturation in long exposures. Even when survey fields sometimes do inevitably include bright stars (e.g. VVDS), galaxies near these stars are rarely selected for follow-up spectroscopy, hampering the use of AO. For deep fields, and in many other cases, it would be a significant gain if LGS-AO without a tip-tilt star allowed one to achieve a resolution better than the seeing limit.

A quantitative example of the advantage – in terms of number of sources accessible – has been given by Mannucci (2007). He selected sources from the survey of about 1000 Lyman Break Galaxies, based on whether there is a nearby star and if the source is at a redshift conducive to near-infrared observations. The result he finds is that none can be observed profitably with NGS-AO, and only about 10 with LGS-AO. But by dispensing with the tip-tilt star, one can find nearly 50 suitable targets.

This increase in number of sources available is actually the same effect that can be seen in figures of the sky coverage as a function of Strehl ratio for NGS- and LGS-AO, which have been published in numerous places. Such figures demonstrate first that sky coverage with LGS-AO is much higher than with NGS-AO. But they also show that the sky coverage increases as the acceptable/achievable Strehl decreases. Fortunately, for LGS-AO a low Strehl is not necessarily bad. The reason is that the flux in the core of the PSF (which depends on the high order correction from the LGS) is independent of the tip-tilt star. Instead, the low Strehl is due to a slight broadening of the PSF core, although its FWHM remains much better than the seeing limit. The cause is the residual image motion (tip-tilt jitter) from the natural tip-tilt star which may be faint and far off-axis. Using a tip-tilt star that is faint and/or far off-axis is nearly the same as not using one at all. The big advantage of dispensing with tip-tilt completely is, of course, that one has fully 100% sky coverage.

## Performance Simulations and Measurements

Based on estimates of atmospheric tip-tilt (direction of arrival statistics) in typical conditions, one might expect image motion – and hence the resultant resolution – to be of order 0.4" independent of wavelength. However, simulations suggest one should do rather better than this. The simulations presented here have been set-up specifically for the 7x7 lenslet array of NAOS on the VLT, assuming 0.8" seeing (at 500nm). The noise level was adjusted so that with full LGS-AO and perfect tip-tilt correction, one achieves about 35% Strehl in the K-band and a FWHM of about 70mas. This corresponds well to the better measurements made with NACO (Kasper 2007). As expected, one then finds that as the number of photons available for tip-tilt correction decreases, so does the predicted Strehl ratio. In the limit of no photons (i.e. no tip-tilt correction), the Strehl is about 11%, corresponding to a FWHM of about 120mas. In both cases, these values correspond well to measurements that have actually been made with NACO (Fig 1). If one considers encircled energy, then the price to pay for dispensing with tip-tilt appears very affordable. The simulations indicate that both with and without tip-tilt, 50% of the flux – twice that for the seeing limited case – remains within a 0.3"x0.3" aperture (Fig 2).

In fact this is not the full story. The simulations do not include wind shake and other vibrational effects that are responsible for a significant amount of jitter, and which might strongly limit the resolution on these scales. Without a tip-tilt star these will not be corrected. However, the stabilisation provided by the actuation of the secondary mirror of the VLT, which can run with a measurement frequency of up to 30Hz, acts as a 'hidden' tip-tilt correction which is going on all the time. As such it takes care of most of these effects without limiting the sky coverage. It allows one to reach resolutions that are better than otherwise expected, and close to the predictions above.

**Astronomical Applications**

There are two adaptive optics instruments on UT4 which are able to make use of the Laser Guide Star Facility. The near infrared integral field spectrometer SINFONI, and the imaging spectrometer NACO. This section describes observations made with both of these instruments that demonstrate a few science cases where the seeing enhancement afforded by LGS-AO without tip-tilt is beneficial; and outlines some of the reasons why one might want to consider using it.

One object was already observed using this mode accidentally during the early phases of commissioning, when the tip-tilt loop failed to close but the integration continued (a feature which was quickly corrected). It was the ultraluminous infrared galaxy IRAS 11095-0238 (Fig 3), which was easily resolved into 2 close nuclei separated by only 0.53". This corresponds to 1.8kpc at the galaxy's redshift of z=0.107, indicating that these two nuclei are in the final stages of merging. Earlier optical observations with HST (Bushouse et al. 2002) had also resolved the two bright spots, but it was not clear whether it was instead a single nucleus crossed by a dust lane. These NACO K-band adaptive optics data rule out that possibility.

A young star cluster in NGC1313 was also observed during LGS commissioning. NGC1313 is an unusual isolated galaxy which nevertheless appears to have undergone an interaction. Furthermore, the gas – and hence star formation – is in the outer parts rather than the nucleus (see ESO press release 43/06). One of the star clusters in the north of the galaxy is actively forming stars, and contains in excess of 20 very massive young Wolf-Rayet stars (Crowther & Hadfield, 2007). This cluster was observed in the K-band with NACO and LGS-AO, using a tip-tilt star 50" away. The data reveals numerous individual stars, but the PSF is rather large, 0.3" FWHM, and also appears slightly elongated in the direction of the tip-tilt star (Fig 4). If this is due to the effects of tip-tilt anisoplanatism, then the observations might actually have better been done without tip-tilt.

Arp 220 is a prototypical merger system in which the progenitor nuclei are separated by only 0.9" (400pc), and hence are in the final stages of coalescing (see ESO press release 27/07). This is a very difficult target for adaptive optics because there are no compact sources from which the tip-tilt can be measured; even the optical light from the galaxy itself is extremely diffuse (in stark contrast to the infrared light). Thus the LGS-AO observations are in effect performed without tip-tilt. Nevertheless the resolution achieved is comparable to that from HST at the same wavelength (Fig 5), about 0.15-0.20". However, the SINFONI data are much richer due to the spectroscopic information, and are able to measure the very different kinematics and morphologies of the stars and gas (Davies et al., *in prep*).

Although one might expect it to be easy to find tip-tilt stars for galactic targets, there are on the contrary many such sources where this is not possible. One very clear example is the Butterfly star system, which consists of an edge-on circumstellar disk around a young low-mass T Tauri star. Scattered light images from the Hubble Space Telescope have shown that the width of the dust lane (i.e. the apparent vertical extension of the disk) decreases dramatically at longer wavelengths (Wolf et al. 2003). This wavelength dependence allows one to investigate the vertical structure of the disk as well as the dust grain properties. At longer wavelengths, the reduction in dust opacity makes it possible to probe deeper layers of the disk. As a result, combining J-band to M-band data enables one to constrain the grain growth processes and the settling of dust grains towards the mid-plane of the circumstellar disk, which are key processes in the early stages of planet formation. The optically brightest star within 60" of the Butterfly Star has R~19.3; and the nearest star bright enough for tip-tilt with R~14.8 is 90" away. Hence one has to use LGS-AO without tip-tilt. However, for L- and M-band observations, individual integrations are rather short and so one can recover at least some of the lost resolution by performing a shift-and-add combination afterwards (Roccatagliata et al., *in prep*).

The final example we give here for LGS-AO without tip-tilt is that of high redshift galaxies, detailed observations of which are a key to understanding galaxy formation and evolution. For such applications, diffraction-limited observations are impracticable because of the faintness of distant galaxies. However, the gain of enhanced resolution that can be provided by LGS-AO without tip-tilt is very substantial given the typical angular sizes of high redshift galaxies of ~1" or smaller. The two z~2 star-forming galaxies shown in Fig. 6 were observed as part of the "SINS" survey (Förster Schreiber et al. 2006a, 2006b, and in prep.), and were selected from the wide-area imaging survey of the "Deep3a" field (Kong et al. 2006). BzK-15504 was observed with both SINFONI using NGS-AO to map the Hα line emission (Genzel et al. 2006, ESO press release 31/06) and also with NACO using full LGS-AO to measure the stellar continuum. The emission is extremely faint: BzK-15504 has a K-band magnitude of 19.2 integrated over about 1". Thus, in terms of sensitivity requirements at the diffraction limit, detecting it is comparable to detecting a point source with K>25. As a result, despite 2 hours integration with NACO using the largest (54mas) pixel scale, the data still had to be smoothed to 0.2" in order to reach sufficient signal-to-noise – as was done for similar reasons to the data from the 6 hour integration on SINFONI using the intermediate (100mas) pixel scale. The effective resolution of 0.2", which can be achieved without tip-tilt, is nevertheless of extreme scientific value, corresponding to a physical scale as small as 1.6 kpc at the redshift of BzK-15504. For the SINFONI observations of the Hα line emission of BzK-6004 (Shapiro et al. 2008; Genzel et al. *in prep*), the largest pixel scale was used to maximize the observing efficiency and signal-to-noise. This gives an 8" field of view within which it is possible to dither, and so no additional sky frames are needed. In this case, the spatial resolution is then limited by the pixel scale rather than by the lack of a suitable tip-tilt star. Hence LGS-AO is used simply to enhance the resolution to 0.4" rather than reach the diffraction limit. Thus, the lack of a tip-tilt star had no direct impact on the observations.

**Prospects for Future Observations**

There are many astronomical instances where there are no suitable tip-tilt stars for laser guide star adaptive optics, but for which improved spatial resolution can bring immense benefits to the scientific analysis. Simulations and observations have shown that LGS-AO without tip-tilt stars does work on the VLT, and that the typical K-band resolution that can be achieved in good conditions appears to be around 0.2" with 100% sky coverage. At the time of writing, ESO's adaptive optics team is hoping to obtain additional technical data to evaluate this seeing enhancement mode further. In closing, we are glad to report that this mode, although not yet fully commissioned, has been implemented on SINFONI; and that there are plans to implement it also on NACO. Although the LGS can only be used in good seeing conditions – which inevitably limits the time it can be available – it is likely that LGS-AO without tip-tilt may soon become available to the community

**FIGURES**

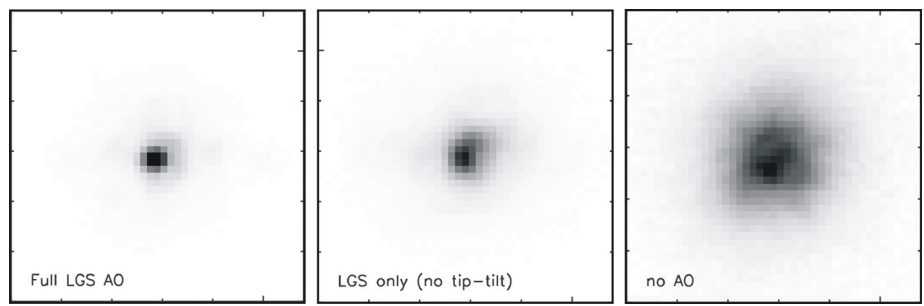

**Figure 1** – NAOS performance measured on CONICA in the K-band for full LGS-AO (left), and without tip-tilt (centre). The seeing limited PSF (right) is shown for reference. With full LGS-AO, the Strehl ratio was 22% and the FWHM ~85mas. Without tip-tilt, the Strehl ratio was reduced to 10% and the FWHM increased to 130mas. For comparison, the K-band FWHM without AO is 300mas.

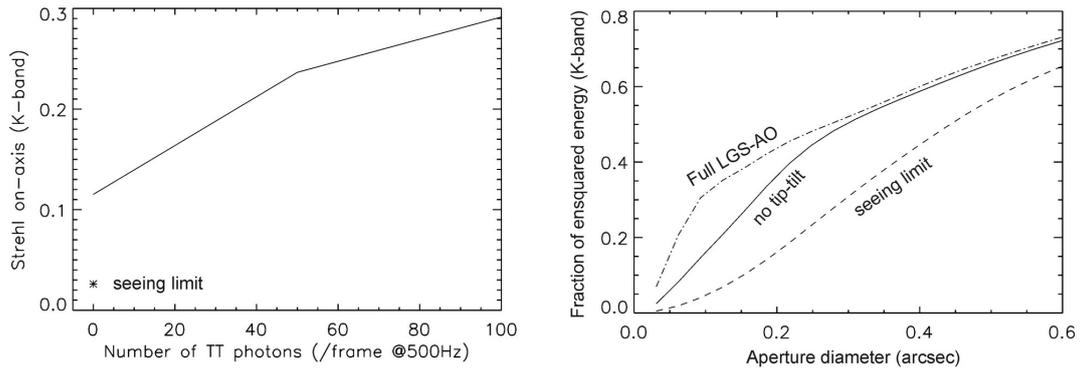

**Figure 2** – simulated performance for LGS-AO with and without tip-tilt. Left: predicted K-band Strehl as a function of the number of tip-tilt photons, tending towards 35% for full LGS-AO and 11% with no tip-tilt. Right: the ensquared energy as a function of aperture size shows that tip-tilt has no impact on the flux measured in a box of side 0.3" or more, and that this flux is still significantly higher than for the seeing limited case.

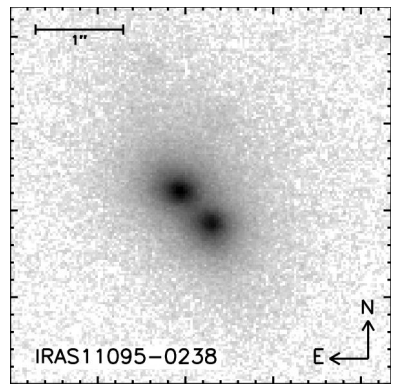

**Figure 3** – K-band image of the ULIRG IRAS 11095-0238 taken using LGS-AO without tip-tilt during LGSF commissioning. The two progenitor nuclei are clearly resolved to have a separation of 0.53".

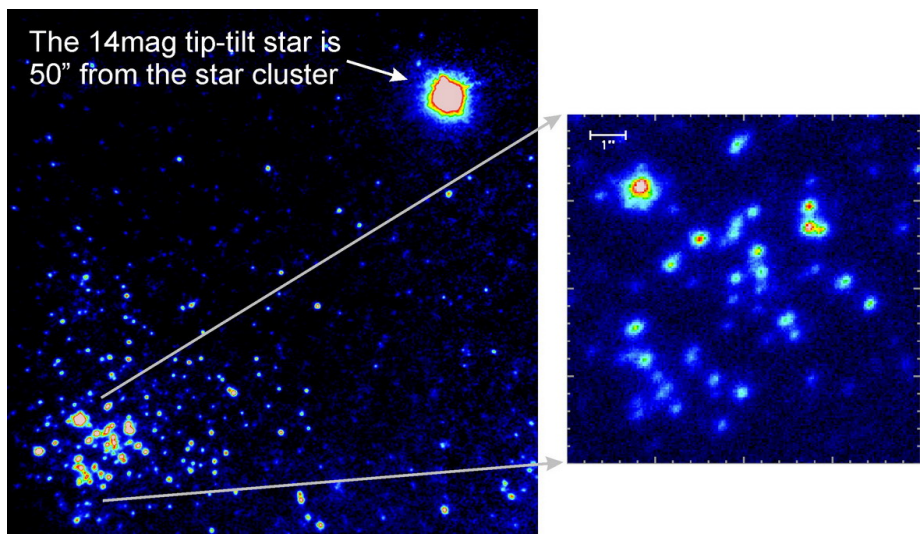

**Figure 4** – NACO K-band image of a star cluster in NGC1313 taken with full LGS-AO during LGSF commissioning. Left: the full image shows the star cluster (lower left) 50" from the bright tip-tilt star (top right). Right: a close-up of the star cluster reveals that the PSFs have a FWHM of about 0.3" and are elongated towards the tip-tilt star.

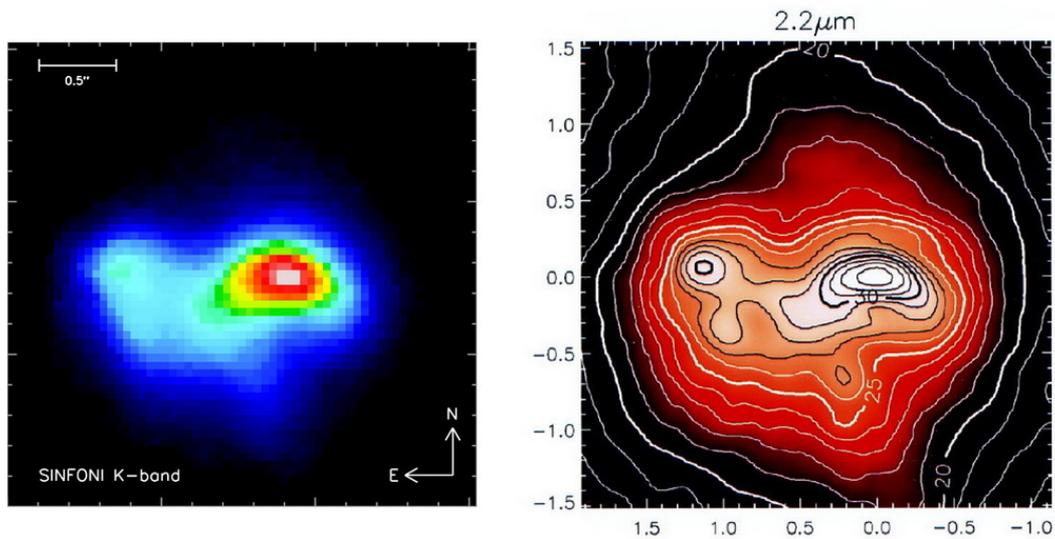

**Figure 5** – 3"x3" K-band images of the prototypical merger galaxy Arp220. Left: SINFONI with LGS-AO but effectively no tip-tilt (Davies et al. in prep). Right: HST NICMOS (from Scoville et al. 2000). Both images show the same level of detail (resolution 0.15-0.2"), but the SINFONI dataset is much richer due to the spectroscopic information.

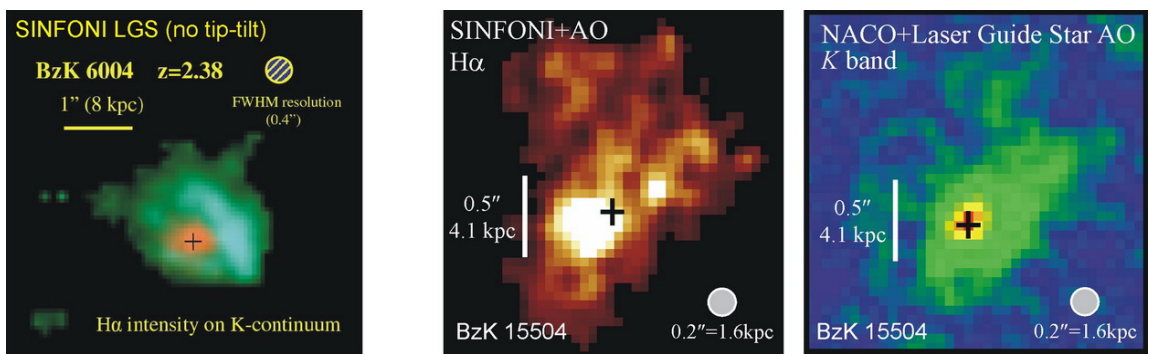

**Figure 6** – High redshift (z=2.4) galaxies observed with SINFONI and NACO. Left panel: BzK-6004 (Shapiro et al. 2008; Genzel et al. *in prep*) has no tip-tilt star; LGS-AO was used with the large pixel scale to enhance the resolution while still allowing dithering within the field to maintain observing efficiency. Centre and right panels: BzK-15504 (Genzel et al. 2006, ESO press release 31/06) was observed using NGS-AO for SINFONI and full LGS-AO for NACO. But because the targets are so faint, both datasets had to be smoothed to 0.2" resolution in order to reach sufficient signal-to-noise.